\documentclass[pdftex,twocolumn,epjc3]{svjour3}          

\RequirePackage[T1]{fontenc}

\smartqed  

\pdfoutput=1 

\RequirePackage{graphicx}
\RequirePackage{mathptmx}      
\RequirePackage{flushend}
\RequirePackage[numbers,sort&compress]{natbib}
\RequirePackage[colorlinks,citecolor=blue,urlcolor=blue,linkcolor=blue]{hyperref}

\newcommand {\epem}{$e^+e^-$}     

\journalname{Eur. Phys. J. C}

\begin{document}

\title{Study the effect of beam energy spread and detector resolution on the search for Higgs boson decays to invisible particles at a future e$^+$e$^-$ circular collider}


\author{
Olmo Cerri\thanksref{addr1}
\and
Michele de Gruttola\thanksref{e1,addr2}
\and 
Maurizio Pierini\thanksref{addr2}
\and 
Alessandro Podo\thanksref{addr1}
\and 
Gigi Rolandi\thanksref{addr1,addr2}
 }


\thankstext{e1}{e-mail:michele.de.gruttola@cern.ch}

\institute{
Scuola Normale Superiore, Pisa, Italy \label{addr1}
 \and
 CERN, Geneva, Switzerland\label{addr2}
 }

\date{Received: date / Accepted: date}

\maketitle

\begin{abstract}
We study the expected sensitivity to measure the branching ratio of Higgs boson decays to invisible particles at a future circular \epem collider (FCC-ee) in the process $e^+e^-\to HZ$ with $Z\to \ell^+\ell^-$ ($\ell=e$ or $\mu$)  using an integrated luminosity of 3.5 ab$^{-1}$ at a center-of-mass energy $\sqrt{s}=240$ GeV. 
The impact of the energy spread of the FCC-ee beam and of the resolution in the reconstruction of the leptons is discussed.
The minimum branching ratio for a $5\sigma$ observation after 3.5ab$^{-1}$ of data taking is  $1.7\pm 0.1\%(stat+syst) $. The branching ratio exclusion limit at 95\% CL is  $0.63
 \pm 0.22\%((stat+syst))$.
\end{abstract}



\section{Introduction}

The absence of any evidence for new physics at the LHC has turned our description of the electroweak scale even more puzzling. The discovery of the Higgs boson with a mass of 125 GeV by ATLAS and CMS~\cite{ATLASHIGGS,CMSHIGGS} increased the urgency to understand the hierarchy problem. The nature of dark matter, the origin of the baryon asymmetry in the Universe, the understanding of the very small neutrino masses are big questions, still missing an answer. These answers cannot be found within the Standard Model (SM).

Some of these open questions could be answered by a new generation of particle colliders as the Future Circular Colliders (FCC)~\cite{FCC}, a set of proposals for a proton-proton, \epem, and e-proton colliders to be hosted in a 100 km tunnel in the CERN area. 

We concentrate on the \epem-collider option (FCC-ee)~\cite{FCCeeweb,TLEP} and we explore its sensitivity to the decay of the Higgs boson to invisible particles. The basic design of the FCC-ee consists in a top-up booster and separate e$^+$ and e$^-$ beams, allowing to reach very large luminosities. The present baseline figure for FCC-ee luminosity\cite{Zimmermann} at $\sqrt{s}=240$~GeV is 1~ab$^{-1}$ per year with two interaction points and the design target figure is 3.5~ab$^{-1}$ per year with four interaction points.


A coupling of the Higgs boson ($H$) to non SM invisible particles is predicted in many extensions of the SM, as for instance in Higgs-portal model~\cite{HiggsPortal} of Dark Matter (DM). In this scenario, one could explain why DM particles were not yet detected in underground experiments, while easily accommodating the experimental picture emerging from the Run-I LHC data.

\begin{figure}[htb] 
\begin{center} 


\includegraphics[width=0.5\textwidth]{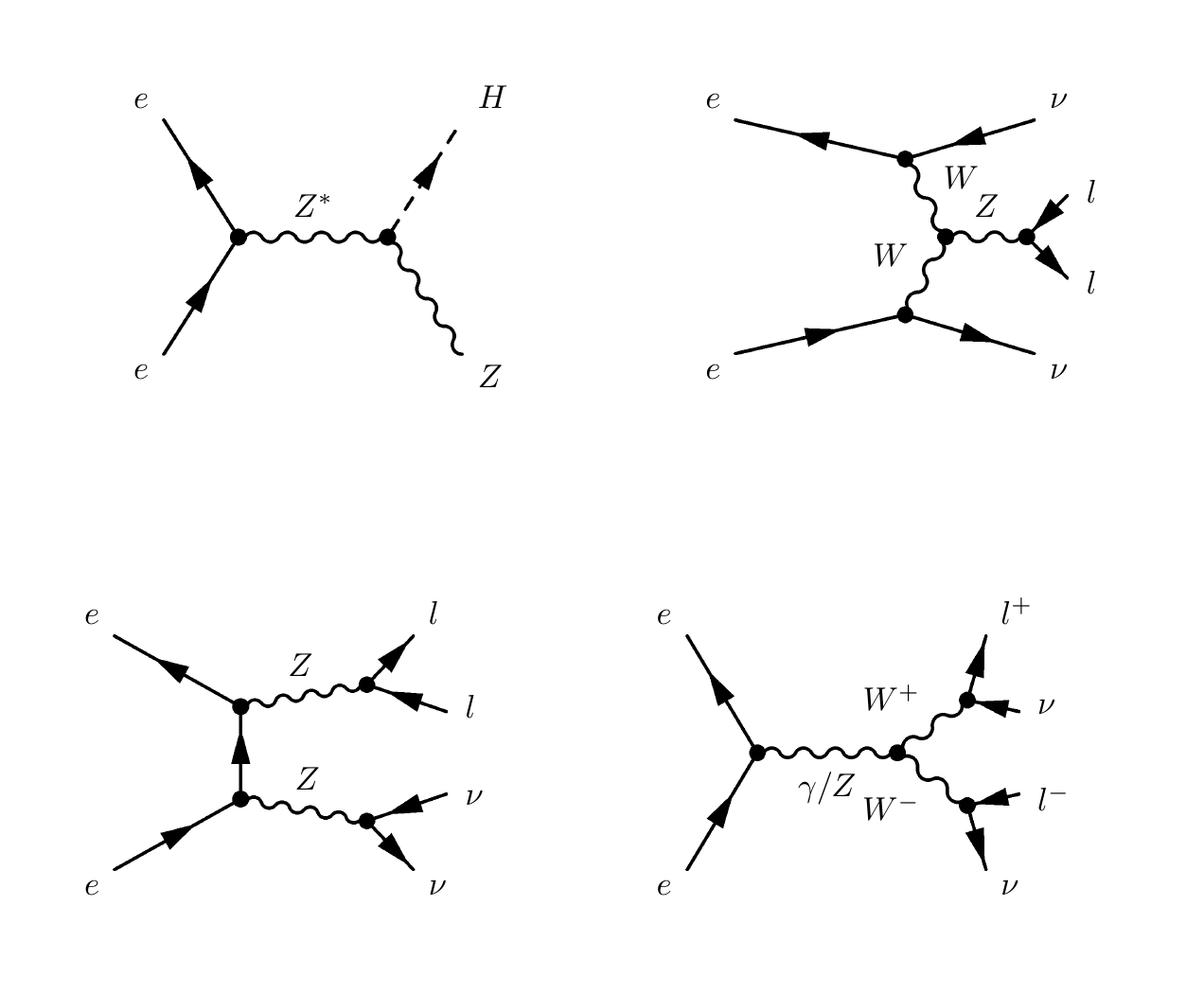} 

\caption{Feynman diagrams for the main production mechanisms for: (top left) $ZH$ 
signal production; (top right) $Z \nu \bar{\nu}$ production; and  (bottom) $ZZ$ and $WW$ 
production.\label{fig:FeynDiag}}
\end{center}
\end{figure}

At the FCC-ee, $H$ bosons could be copiously produced in association to $Z$ bosons (see Fig.~\ref{fig:FeynDiag}), operating the collider above the $m_Z$+$m_H$ energy threshold, where $m_Z$ and $m_H$ are the $Z$ and $H$ boson masses. At $\sqrt{s}=240$~GeV, the largest contribution to the $H$ production cross section is given by Higgsstrahlung process $e^+e^-\to HZ$ whose cross section at this energy is 201 fb, 
as estimated with PYTHIA8~\cite{PYTHIA8}. 

Invisible $H$ decays result in a  mono-$Z$ signature, in which a $Z$ boson is detected in events with no visible particle balancing its momentum.  These events can be identified reconstructing the $Z$ boson and searching for an  excess at 125 GeV in the distribution of the event missing mass, recoiling to the reconstructed Z boson.

In the clean environment provided by the FCC-ee, one can tag $HZ$ events through any decay of the $Z$ boson to visible particles. In this study, we concentrate on $Z\to e^+e^-$ and $Z\to \mu^+\mu^-$ final states . Given the expected good resolution for muon and electron momentum measurements, these final states are characterized by the narrowest possible peak in the missing-mass distribution of signal events. 
We will show that the sensitivity of this analysis depends on the momentum resolution and on the beam energy spread giving useful information for the design of the detector and of the accelerator.


Within the SM $H$ bosons can decay to invisible final states through a $ZZ^*$ decay with $Z^{(*)} \to \nu \bar \nu$. The  Branching Ratio (BR) of the full decay chain $H\to ZZ^*\to  2\nu 2 \bar \nu$ is  $\approx 0.1\%$. This figure is small compared to the sensitivity of the analysis discussed in this paper. 

The main SM backgrounds originate from the production of  boson pairs $e^+e^-\to WW(ZZ)\to \ell^+\ell^-\nu \bar \nu$ with the production mechanisms shown in Fig.~\ref{fig:FeynDiag}, which have a cross section times BR in leptons and neutrinos of 370 fb and 36 fb respectively.  
$ZZ$ production mimics the $H \to inv$  $Z \to \ell^+\ell^-$ signal  events when one $Z$ decays to leptons and the other to neutrinos.
These events are characterized by a peak in the missing mass distribution at the $Z$ pole, with a tail to larger values originating from initial state radiation (ISR) of a photon close to the beam axis. 
Opposite charge, same flavor leptons originating from independent $W$ decays in $WW$ events may have an invariant mass close to the $Z$ pole and thus be miss-tagged as a real $Z$ boson recoiling against invisible particles. Due to the large $WW$ cross section, this background is not negligible. Additional processes like $e^+e^- \to Z\nu\nu$ (see Fig.~\ref{fig:FeynDiag}) are found to be negligible~\cite{TLEP}, given the small cross section. Moreover, we verified that events with radiative return to the Z peak are completely rejected by the selection criteria discussed in section \ref{sec:EvtSec}, and that  we can safely neglect  any sources of acoplanar leptons in  $\gamma \gamma$  processes.

The mono-$Z$ and other signatures have been already explored at the LHC, resulting in an upper limit on the H boson invisible branching ratio of 25\%\cite{ATLASHiggstoInv,CMSHiggstoInv}. 
Interesting constraints are derived on the DM-nucleon scattering cross section in Higgs portal models. Assuming the total $H$ width to agree with the SM prediction, a more stringent bound on $\Gamma_{inv}$ can be put from a global analysis of the $H$ couplings to visible SM particles~\cite{GiaStrumia}.

Sensitivity studies of the invisible Higgs boson branching ratio measurement at future \epem colliders exploiting the Higgsstrahlung process and the missing mass technique have been performed in the context of the International Linear Collider~\cite{ILCPROJECT} , of a 50-70 km long circular electron positron collider (CEPC)~\cite{CEPC} proposed by the Chinese high energy physics community and also in a first look at the physics case of FCC-ee~\cite{TLEP}. They~\cite{HANLIU,CEPCTDR} show that significantly better sensitivity can be obtained using also the channel $HZ$ with $Z$ decaying into hadrons in spite of the lower missing mass resolution because of its larger statistics. 


This paper is organized as follows: Section~\ref{sec:EvtGen} describes the relevant physics process and the procedure to generate the corresponding Monte Carlo (MC) samples; Section~\ref{sec:Detector} discusses the approximations used to incorporate in the analysis the resolution and efficiency effects of a realistic detector simulation. The events selection and the analysis strategy and results are described in Section~\ref{sec:EvtSec} and Section~\ref{sec:results}, respectively. 

\section{Event Generation}
\label{sec:EvtGen}

Signal and background samples are produced using the \\ PYTHIA8~\cite{PYTHIA8} MC leading-order event generator.  We generate $WW$, $ZZ$ and $HZ$ events in which the $W$ and $Z$ bosons are forced to decay in leptonic channel ($e$, $\mu$ and $\tau$). No additional generator-level filter is applied. The possibility of exploiting $Z \to \tau^+ \tau^-$ decays to increase the signal yield is not investigated, given the worse resolution for the missing-mass peak. However this decay mode provides a further source of non-peaking background, when the two $\tau$ leptons decay to a pair of same-flavor and opposite-sign electrons or muons.

$H$ bosons are forced to a decay to a pair of neutralinos $\tilde{\chi}^{0}_{1}$ with mass $m_\chi = 5$~GeV. 
The use of this specific benchmark for invisible particle does not limit the generality of our results, as long as the condition $2 m_\chi< m_H$ is fulfilled. 

\section{Detector Simulation}
\label{sec:Detector}

One of the goals of this study is to define criteria to be used in the design of a detector for FCC-ee. The comparison of  the sensitivities reachable at FCC-ee using 
detector concepts with different resolutions gives useful information. In order for this study to be performed in a realistic condition, the beam-energy spread expected at the FCC-ee ($0.17\%$ on single beam, $0.12\%$ on the center of mass energy) is included when simulating the \epem collisions.

 
In this study detector effects are simulated using the Delphes 3.2.0~\cite{DELPHES} parametric simulation with different conditions. As conservative design we have chosen the CMS detector parametrized with\cite{CMSDELPHES} and the relevant distribution of the missing mass to the lepton pair is compared for validation with a similar study~\cite{LEP3} performed with full simulation of this detector. As a more performing design, we have used the parametrization~\cite{ILDDELPHES} of one the two ILC detector detector designs, being aware that this is a crude approximation since the linear collider environment differs in an significant way from the circular collider one with implications on detector parameters like cooling which can increase the detector mass.




The reference system used in this analysis has the origin at the nominal collision point, the z axis along the electron direction and the x axis toward the center of the collider. The polar angle $\theta$ is defined with respect to the z positive axis. The projection of the momentum on the plane perpendicular to the beams is $p_T$ and the pseudorapidity $\eta$ is $\eta=-\ln(\tan(\theta/2))$.

The major differences between the two detector parametrization are listed below:
\begin{itemize}
\item Solenoid:
        \begin{itemize}
          \item Magnetic field strength: $B_Z$: 3.5~T at ILD, 3.8~T at CMS.
          \item  Tracking radius: 1.8~m at ILD, 1.29~m at CMS.
          \item Half length of field coverage: 2.4~m at ILD, 3.0~m at CMS.
        \end{itemize}
\item Tracking efficiency:
        \begin{itemize}
          \item ILD: 99\% for particles with $p_T>100$~MeV and $|\eta|<2.4$, including muons and electrons.
          \item CMS: 95\% for particles with $p_T>100$~MeV and $|\eta|<2.5$, including muons and electrons.
        \end{itemize}
\item Muon momentum resolution:
        \begin{itemize}
          \item ILD: $\frac{\Delta P}{P} = 0.1\% + \frac{P_T}{10^5 GeV}$ for $|\eta|<1$ and 10 times higher for $|\eta|$ up to 2.4. 
          \item CMS: between $1\%$ and $5\%$. 
        \end{itemize}
\item Electron energy resolution:
        \begin{itemize}
          \item ILD: $\frac{\Delta E}{E}= \frac{16.6\%}{\sqrt{E[GeV]}} + 1.1 \%$.
          \item CMS: $\frac{\Delta E}{E} = \sqrt{E^2*0.007^2 + E*0.07^2 + 0.35^2}$, $E$ in GeV.
        \end{itemize}
\item Particle reconstruction efficiency: 
        \begin{itemize}
            \item  ILD:  99\% for $e, \mu$ and $\gamma$ with $P_T>10$~GeV. 
            \item CMS:  85\%-95\% for the same $p_T$ range.
        \end{itemize}
\end{itemize}

When running the Delphes detector simulation, the Particle Flow (PF) reconstruction option is activated, which produces a list of {\it reconstructed} particles (electrons, muons, photons, charged hadrons and neutral hadrons), to which an event selection is applied. More details on the Delphes implementation of the PF algorithm can be found in Ref.~\cite{DELPHES}.


\section{Event Selection}
\label{sec:EvtSec}

Signal and background events are selected applying the following requirements to the PF particles returned by Delphes:

\begin{itemize}
\item Reject events with photons with $p_T>20$~GeV.
\item Exactly two opposite-charge muons or electrons with $p_T>10$~GeV.
\item At most one reconstructed photon with $p_T>10$~GeV, which could be an ISR or FSR photon.  If present, the photon is considered to be the FSR of one of the two leptons if it is closer than $dR=\sqrt{\delta\eta^2 + \delta\phi^2}<0.4$ to the lepton. Its momentum is added to the di-lepton four-momentum, to reconstruct the $Z$ candidate four-momen\-tum. 
\end{itemize}
Following Ref.~\cite{LEP3}, the following requirements are applied:
\begin{itemize}
\item Angle between leptons in the laboratory frame $\Delta\theta_{ll}>100$ degrees.
\item Acoplanarity angle $\theta_{aco}>10$ degrees. The acoplanarity angle is defined as the angle between the plane containing the lepton momenta and the beam axis, 
\item Transverse momentum of the lepton pair $p_{T}^{ll}>10$~GeV.
\item Longitudinal momentum of the lepton pair $p_{L}^{ll}<50$~GeV.
\end{itemize}
The first and fourth requirements reduce the $ZZ$ background contribution, while the second and third reject events with a radiative return to the $Z$ pole .  A dilepton pair surviving this selection is considered as a $Z$ candidate if its mass is found to be within 4~GeV from $m_Z$.

The full selection efficiency for the signal is  74\%, while for ZZ and WW backgrounds is  36\% and  3\% respectively.

\section{Analysis Strategy and Results}
\label{sec:results}

The main signal-to-background discriminating power comes from the knowledge of the four-momenta of the
colliding leptons.
 This information is used to compute the four momentum of the missing particles in the event, by difference. When all missing particles in an event come from the decay of a mother particle, the invariant mass computed from the missing four momentum {\it resonates} at the value of the mother-particle mass. 
 We compute the event missing mass as:
\begin{equation}
M_{\rm miss} = \sqrt{(\sqrt{s}-E_Z)^2 - |{\vec p}_Z|^2}
\end{equation}
where $(E_Z, {\vec p}_Z)$ is the four-momentum of the $Z$ boson candidate, computed from the sum of the four-momenta of the dilepton pair and, when found, 
an FSR photon. $\sqrt{s}$ is the nominal collision energy. 

Figure~\ref{M_miss_comparison} shows the missing mass distribution where a branching ratio $BR(H \to inv)=100\%$ has been assumed for illustration purposes. One notices the peaking ZZ background with a tail extending in the signal region and the non peaking WW background. This figure has been drawn assuming an integrated  luminosity of 0.5~ab$^{-1}$ for direct comparison with figure 6~a of reference~{\cite{LEP3} done for the same channel with full simulation of the CMS detector. A good  agreement is found on the width of the $H$ and $Z$ peaks validating the Delphes simulation of this simple channel.

\begin{figure}[htb]
\centering
\includegraphics[width=0.5\textwidth]{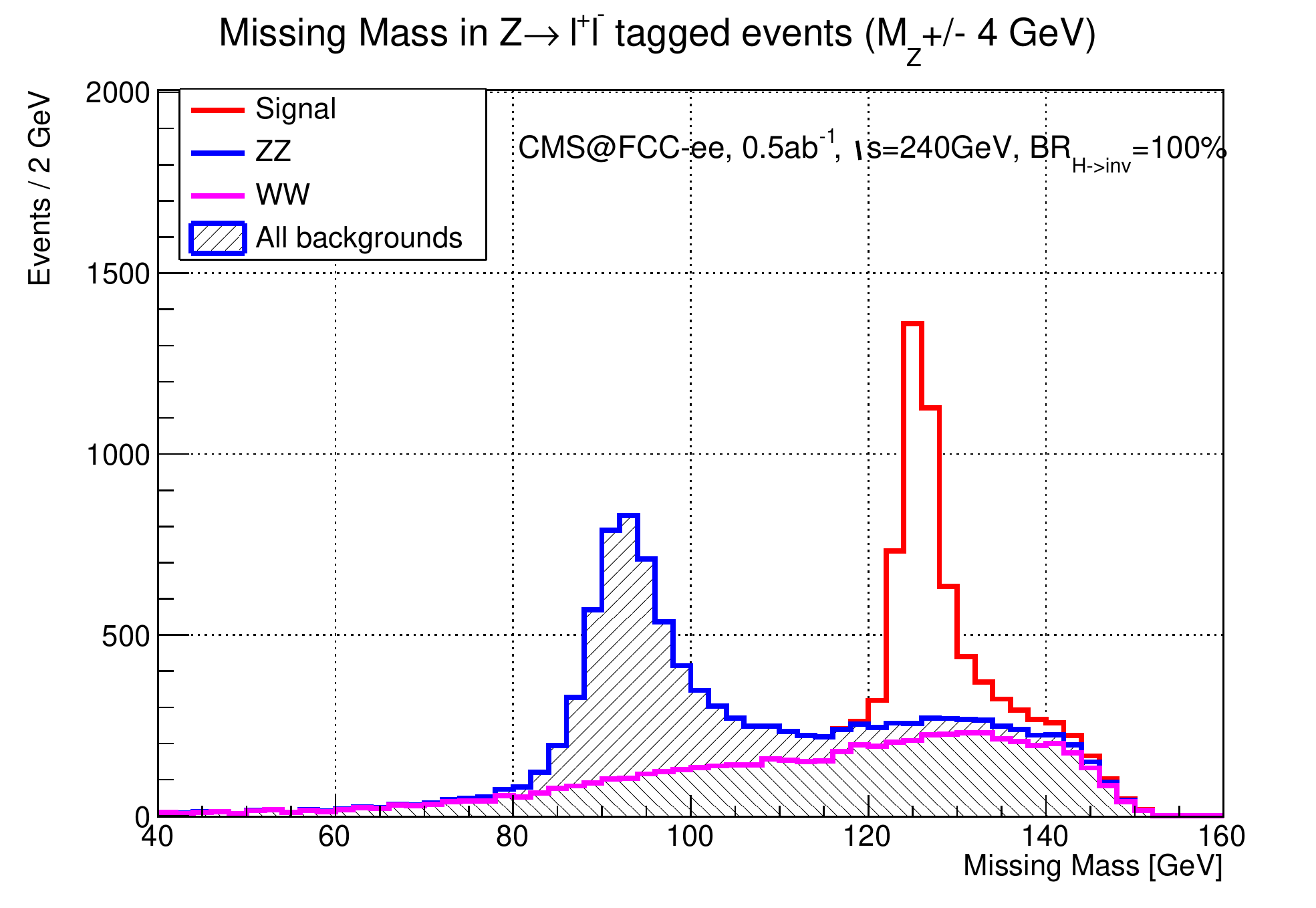} 
\caption{Missing mass distribution from simulation with $BR(H \to inv)=100\%$ and 
the selection requirements described in the text.}
\label{M_miss_comparison}
\end{figure}

The $H \to inv$ signal is extracted from a template fit to the  $M_{\rm miss}$ distribution, using as templates the distributions of the individual processes, derived from MC. 

In an analysis with real data, control samples will be used to validate the agreement between data and MC and/or derive the template distributions. $ZZ \to 4\ell$ and $WW \to e \nu_{e} \mu \nu_{\mu}$ events provide control samples to study the $ZZ$ and $WW$ backgrounds. 
In this work, we don't attempt to simulate the precision that these control-sample studies could reach. Instead, we assume that the uncertainty on the template distributions could be reduced to a negligible level, by using a combination of data control samples and accurate MC simulation.

The analysis performances are quantified generating pseudo datasets with a total yield distributed around the expected yield. The randomization of the total yield is done assuming a Poisson distribution for the total event counting. By running the template fit on each pseudo-experiment, a determination of $BR(H\to inv)$ and the corresponding uncertainties are derived. The exercise is performed as a function of the true value assumed for $BR(H\to inv)$ in generation. In particular, fixing $BR(H\to inv)=0$ in generation, a distribution is derived for the 95\% upper limit on the H invisible branching ratio.

\begin{figure}[htb]
\centering 
\includegraphics[width=0.5\textwidth]{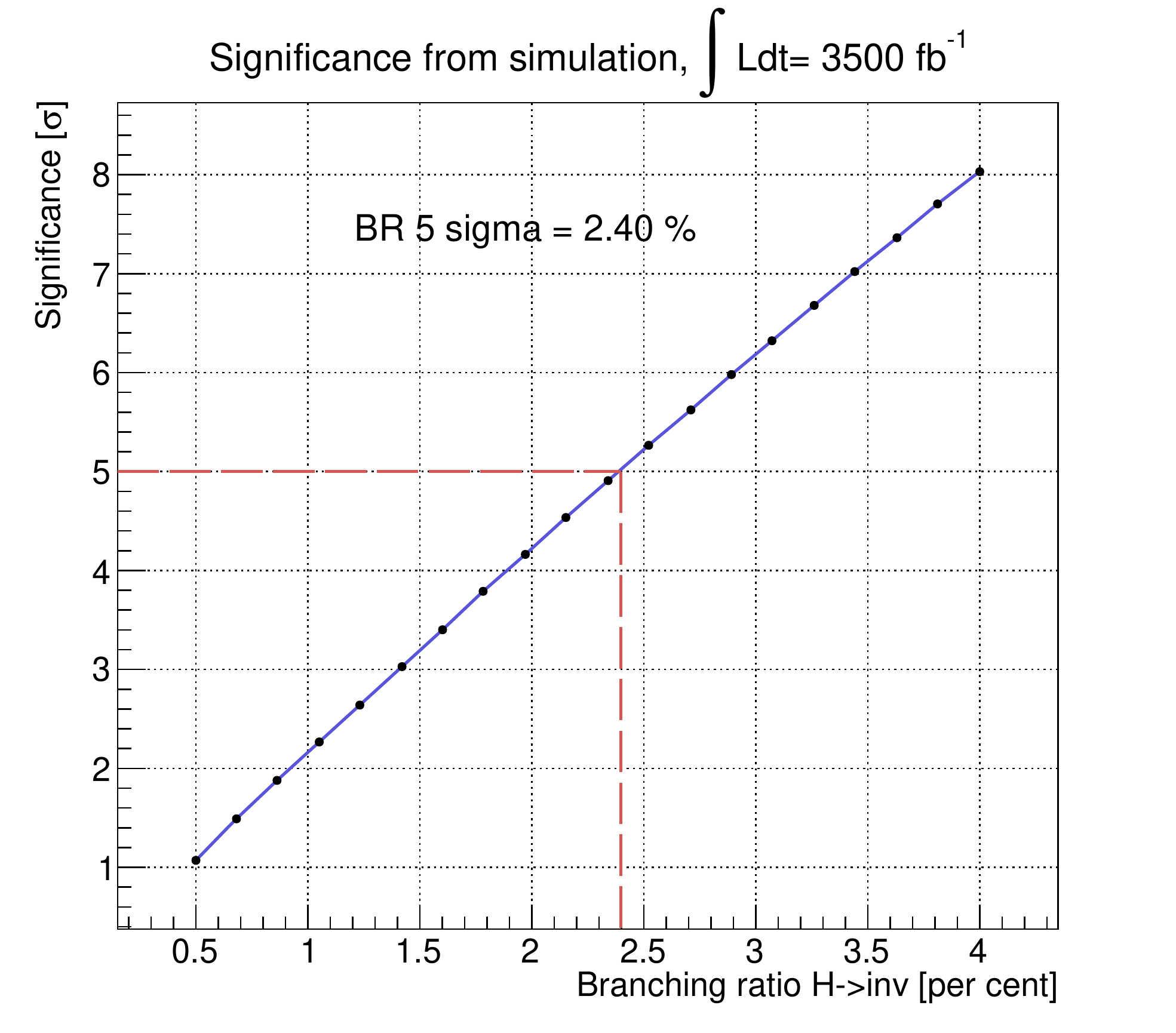}
\includegraphics[width=0.4\textwidth]{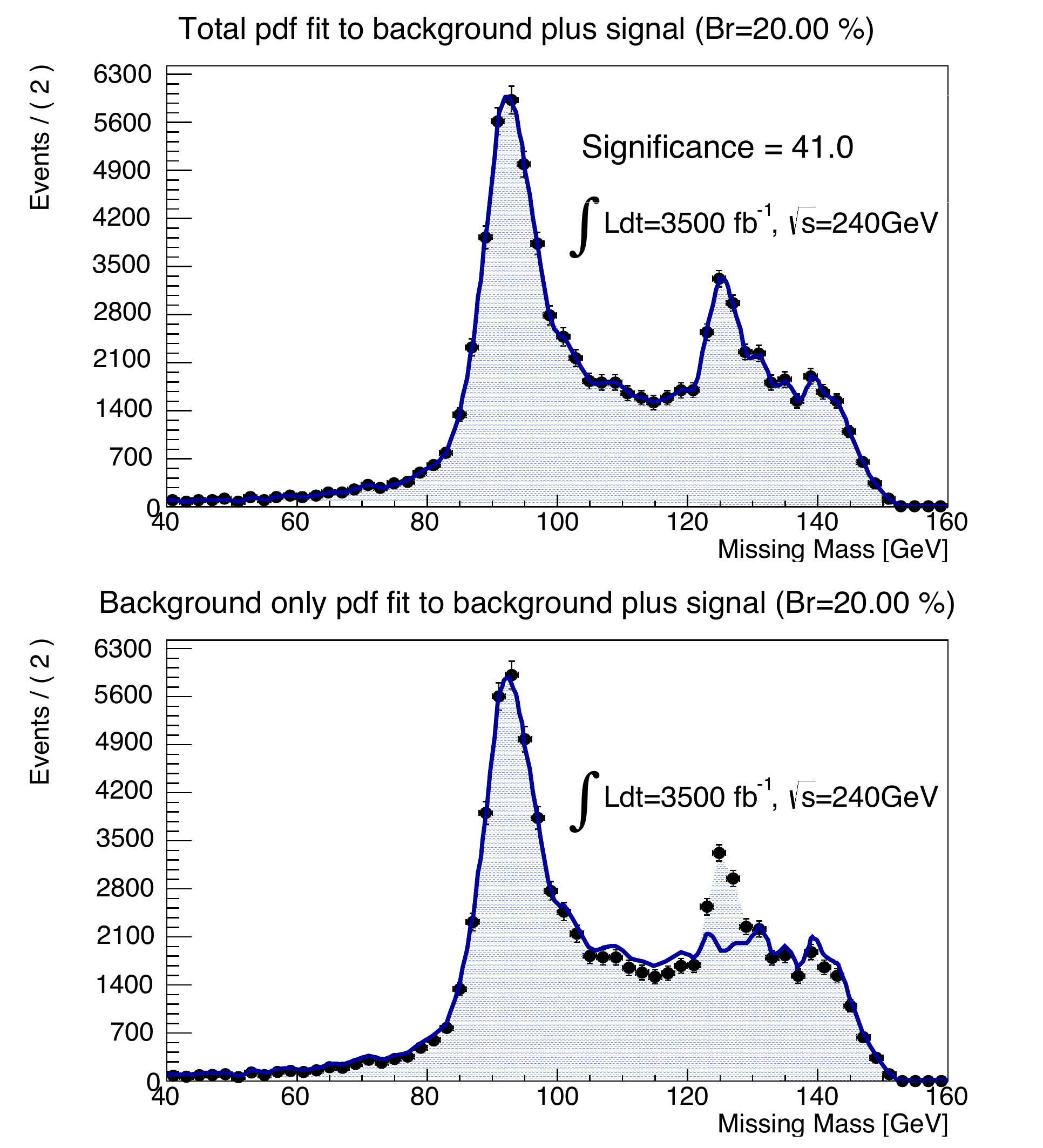}
\caption{Top:  discovery significance as a  function of the $BR(H \to inv)$, derived from template fits to pseudo experiments. 
The 
 line corresponds to the minimum BR value necessary for a $5 \sigma$-significance discovery. 
Middle and bottom: example of a signal+background (middle) and a background-only (bottom) fit for a pseudo data set with 
$BR(H \to inv)=20\%$. The output of these two fits is used to evaluate the significance.}
\label{br_5sigma_expl}
\end{figure}

The sensitivity of a given experimental scenario is evaluated quantifying the minimum discoverable $BR(H \to inv)$ and the corresponding maximum excludable value. For discovery and exclusion we use a reference $5\sigma$ significance and 95\% probability, respectively.

The minimum discoverable $BR(H \to inv)$ is quantified using a set of signal+background pseudo experiments, with a progressive increase of the amount of signal injected at a fixed background amount. Each sample is fit under the signal+background hypothesis, and the likelihood ratio between the best-fit signal and the no-signal hypothesis is used to quantify the significance:
\[ \sigma = \sqrt{-2 \log\frac{\mathcal{L}_{b}}{\mathcal{L}_{s+b}}}~. \]

In the equation, $\mathcal{L}_{b}$ is the maximum likelihood value for a background only fit while $\mathcal{L}_{s+b}$ is the corresponding value for the signal+background hypothesis. Varying the injected $BR(H \to inv)$, we find the lowest BR value corresponding to a $5 \sigma$ significance, as shown in the top plot of Fig.~\ref{br_5sigma_expl}. A Gaussian assumption for the shape of the likelihood is intrinsic in this quantification of the significance. We verified \textit{a posteriori} that such an assumption fairly describes the likelihood distribution for our pseudo experiments.

To evaluate the BR limit at $95\%$ CL, a background-only pseudo experiment is fit many times, for different assumed values of signal yield $N_s$.  The profile likelihood function of $N_s$ is derived from these fits, as shown in Fig.~\ref{likelihood_scan}. An upper limit $N_s^*$on $N_s$ is computed with a Bayesian procedure, integrating the product function. 

\[\int_0^{N_s^*} \mathcal{L}(s+b|N_S) dN_S \equiv 0.95~.\]

The value of $N_s^*$ is translated into an upper limit on the BR normalizing it to the expected number of produced $H$ bosons:

\[BR_{95\%limit}=\frac{N_s^*}{\epsilon L}\]
\noindent

where $\epsilon$ is the selection efficiency (including the $Z \to \ell \ell$ branching fractions) and $L$ is the integrated luminosity.

\begin{figure}[htb] 
\centering 
\includegraphics[width=0.4\textwidth]{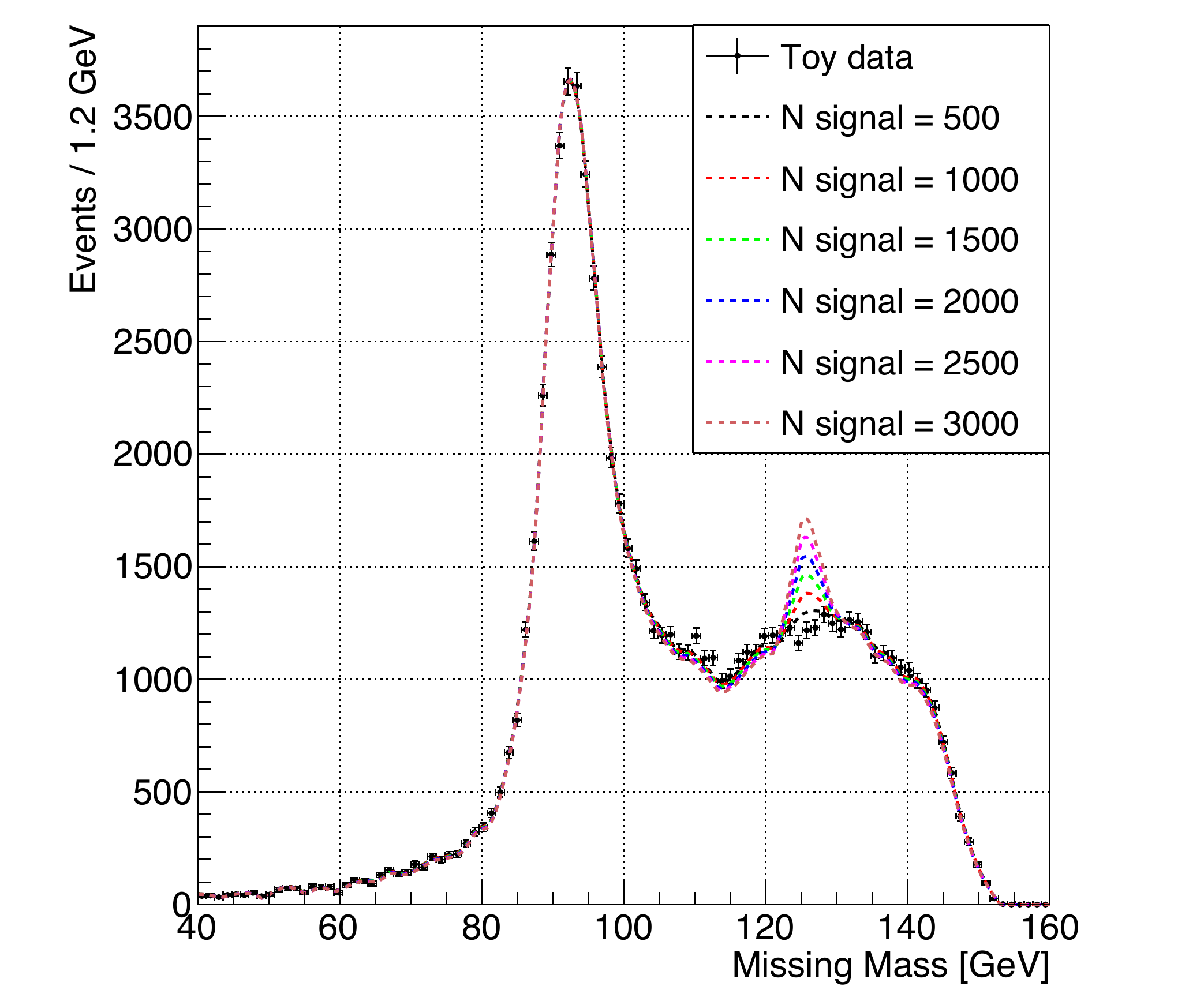} 
\includegraphics[width=0.4\textwidth]{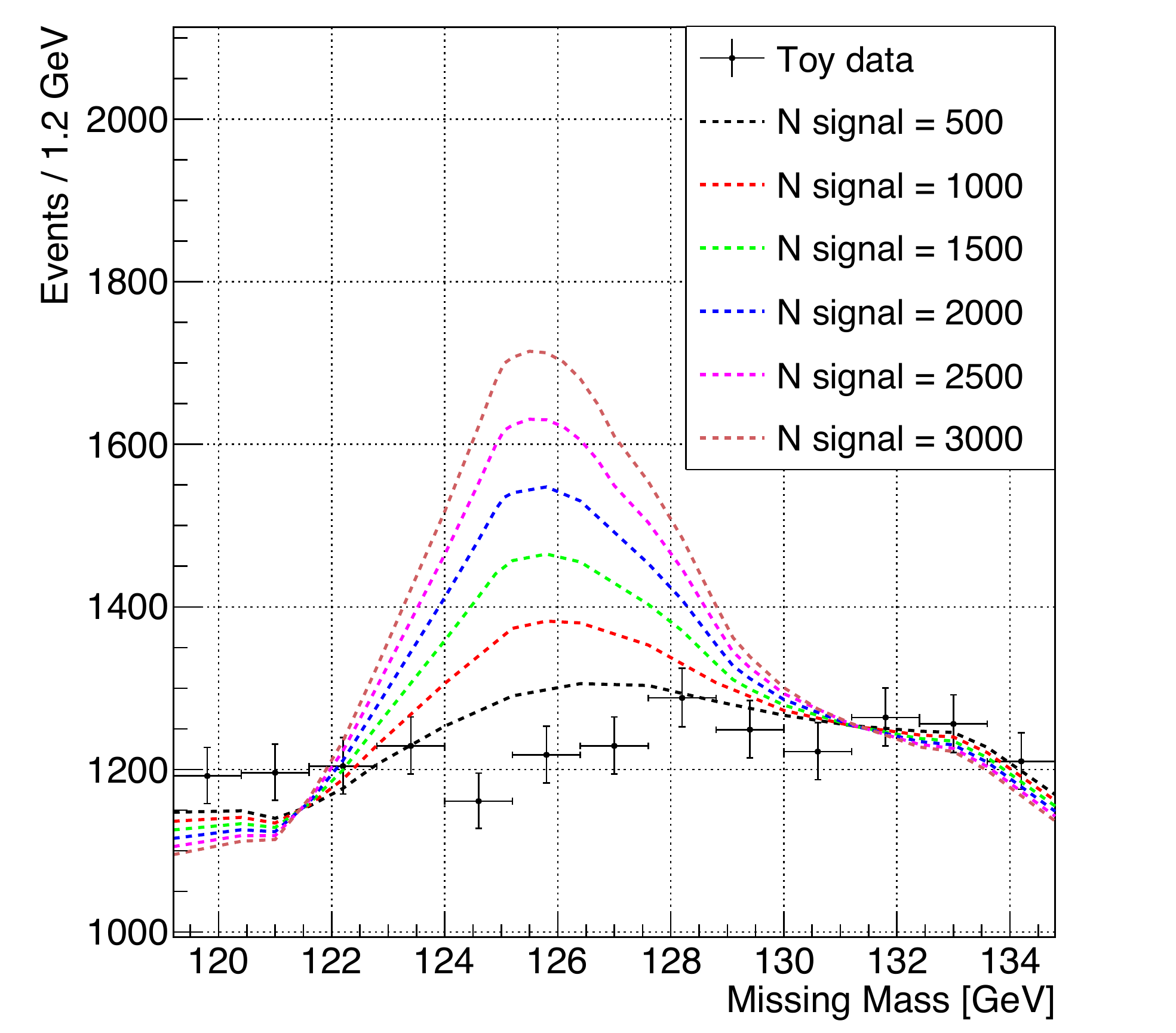} 
\caption{Top: Example of background only toy data set fitted with a pdf signal+background for different values of $N_s$ parameter.
Bottom: a zoom in the $M_{\rm miss} \approx m_H$ region, where the disagreement between the pseudo-data distribution and the signal+background fit increases with 
the increasing assumed value for $N_s$.\label{likelihood_scan}}
\end{figure}

\subsection{Results}
\label{sec:FCCeeres}

We evaluate the expected FCC-ee sensitivity to $BR(H \to inv)$ assuming the ILD-like detector performances.  ILD-like combines excellent tracking capabilities to a finely-grained calorimetry, which makes it a perfect detector for a PF-based reconstruction strategy. The ILD-like performances are compared to those obtained assuming a CMS-like detector.  The CMS-like card set is optimized for hadronic collisions and in particular for a lepton identification designed to suppress the background from {\it fake} lepton candidates from QCD multijet events. If tuned on the topology considered in this work, the lepton identification of the CMS-like detector could be modified increasing the reconstruction efficiency.

We assume an integrated luminosity of 3.5ab$^{-1}$. The $M_{\rm miss}$ distribution is shown in fig.~\ref{ILD_missing_mass}. As expected from the higher resolution and efficiency features, the ILD-like distribution is characterized by $\sim 26\%$ higher efficiency and a narrower peak, both for the $HZ$ signal and the $ZZ$ background. A more accurate comparison between CMS-like and ILD-like detection for this analysis can be found in \ref{appendix_comparison}.

The results for our pseudo-experiment analysis gives 
\[BR_{lim95\%} @ ILD = 0.63\pm (0.22)_{stat} \%\]
\[BR_{5\sigma} @ ILD = 1.7\pm (0.1)_{stat}  \%\] 


\noindent
for the ILD-like detector. Given the high resolution expected for ILD-like, a bin size of 200 MeV has been assumed for the template fit. The systematic uncertainty related to the binning of the templates and the measured energy scale is evaluated varying the bin width by $\pm 50$~MeV and shifting the bin centre up and down by half a bin width. It is found to be negligible.

\begin{figure}[htb]
\centering
\includegraphics[width=0.5\textwidth]{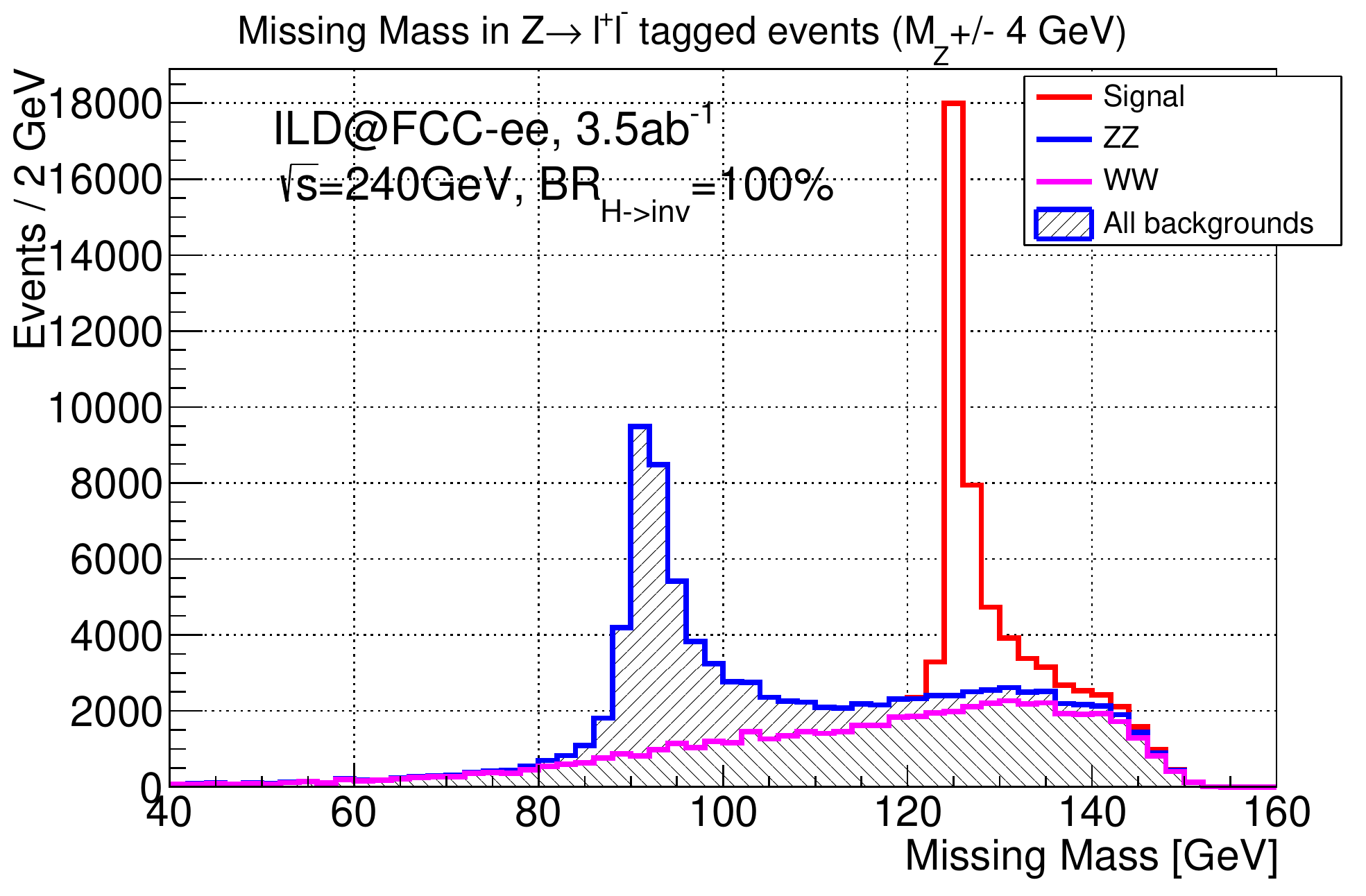}
\caption{Missing mass distribution for a $H\to$ inv $100\%$ branching ratio and 
standard cuts. ILD-like detector was used for this simulation and a luminosity of 
3.5ab$^{-1}$ assumed.\label{ILD_missing_mass}}
\end{figure}
For comparison, the corresponding results for a CMS-like detector and using the same integrated luminosity are:
$$BR_{lim95\%} @ CMS = 0.92 \pm (0.32)_{stat}  \%$$
$$BR_{5\sigma} @ CMS = 2.5 \pm (0.2)_{stat} \%~. $$

These limits on   $BR(H \to inv)$ are  at least an order of magnitude better than the projected precision reachable with the completion of HL-LHC (see tab.2 in~\cite{HLLHC}). They can be compared directly to ILC sensitivity discussed in~\cite{HANLIU} where the analysis performed considering only the decay channels $Z\to e^+e^-$ and $Z\to \mu^+\mu^-$ excludes at 95\% a branching fraction $BR(H \to inv)$ of 3.5\% using 250 fb$^{-1}$ at 240 GeV.

\subsection{Expected sensitivity to Higgs Portal models of Dark Matter}

Under the assumption of SM production cross section, experimental upper limit on the $H\to$inv branching fraction can be used to set a limit on DM-nucleon scattering cross section. This allows to compare the FCC-ee sensitivity to that of direct-detection experiments underground~\cite{Angloher:2011uu,Angle:2011th,Bernabei:2008yi,Savage:2008er,Aalseth:2012if,Agnese:2013rvf,Behnke:2012ys,Akerib:2013tjd}, limited to the specific framework of the Higgs portal model, in which DM particles couple to SM particles only through a $H$ exchange.

The value of $BR(H \to inv)$ is related to the $\Gamma_{Inv}$ by the relation  
\[BR_{inv} = \frac{ \Gamma_{Inv} }{\Gamma_\mathrm{SM} + \Gamma_{Inv}}\] 

\noindent
where $\Gamma_\mathrm{SM}= 4.07$ MeV. Assuming that the DM candidate has a mass $M_\chi < m_{H}/2$, a value for $\Gamma_{Inv}$ can be directly translated into a value for the spin-independent DM-nucleon elastic cross section, according to the following relation (see~\cite{Djouadi2011})~:
\begin{equation}
\label{eq:22}
\sigma^\mathrm{SI}_{\mathrm{S}-\mathrm{N}} =  \frac{4\Gamma_{Inv}}{m_{H}^3v^2\beta} \frac{m_\mathrm{N}^4f_\mathrm{N}^2}{(M_\chi+m_\mathrm{N})^2}
\end{equation}
\noindent

where a scalar (S) DM candidate is assumed (a vector or fermionic case have also been considered but the scalar case is the only one derived from a Lagrangian fully renormalizable, see again~\cite{Djouadi2011}). In equation~\ref{eq:22}, $m_\mathrm{N}=0.939$~GeV is the average nucleon mass, $\sqrt{2}v=246$~GeV is the $H$ vacuum expectation value  and $\beta=\sqrt{1-4M^2_\chi/{m_{H}}^2}$. The quantity $f_\mathrm{N}$ parameterizes the Higgs-nucleon coupling. The nominal values $f_\mathrm{N}=0.326$ is taken from lattice calculations~\cite{fN_LQCD}, while the range found for $f_N$ by the MILC Collaboration~\cite{MILC}, $0.260<f_N<0.629$, is used to estimate a corresponding uncertainty range.  

Following this procedure, the upper limit on $BR(H \to inv)$ discussed in Sec.~\ref{sec:FCCeeres} is translated into a bound on the DM-nucleon cross section.
An improvements of about two orders of magnitude is expected with respect to the current bounds from $H \to inv$ searches at the LHC~\cite{ATLASHiggstoInv,CMSHiggstoInv}, with 3.5 ab$^{-1}$ of FCC-ee run.

Figure~\ref{newDM_limits} shows the comparison of the bound on the DM-nucleon cross section obtained with 3.5 ab$^{-1}$ of FCC-ee run 
with the reach of planned direct detection experiments, such as XENONnT (the upgrade of XENON1T), LZ and DARWIN, which has been elaborated from ~\cite{reviewDM}. 
Note that in the comparison we adopt confidence limits of 90\% as done by the other experiments. The FCC-ee sensitivity would remain competitive for DM masses smaller than 10 GeV.


\begin{figure}[htbp]
 \centering
\includegraphics[width=0.5\textwidth]{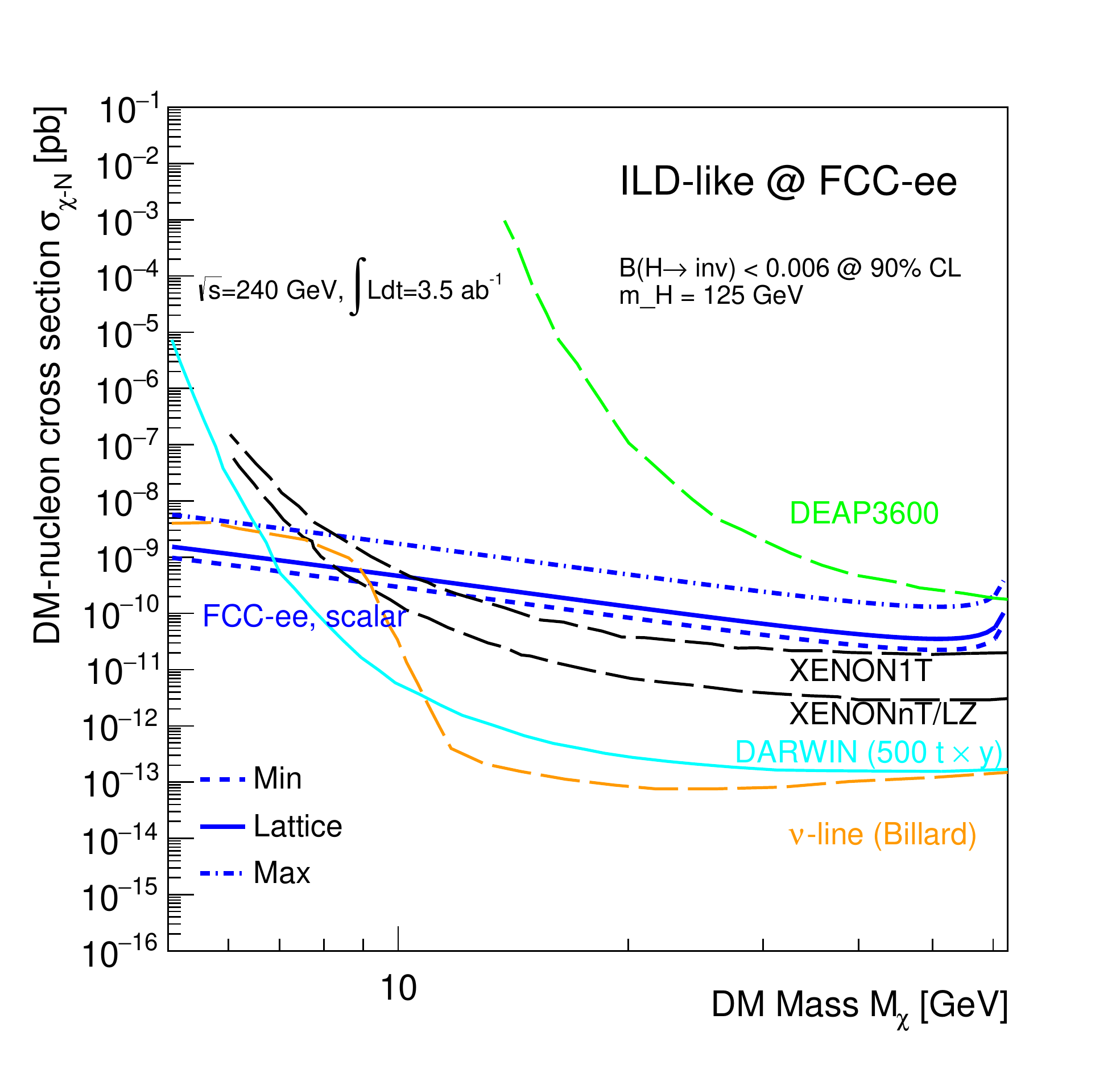}
 \caption{Bounds on DM - nucleon cross section limits that can be set after 3.5 ab$^{-1}$ of data acquisition at FCC-ee with an ILD-like detector compared to the projection of underground direct detection experiments \label{newDM_limits}}
\end{figure}

\section{Summary}
\label{sec:conclusions}

Assuming an integrated luminosity of 3.5ab$^{-1}$ for FCC-ee at $\sqrt{s}=240$~GeV with an ILD-like and a CMS-like detector, we evaluate the discovery and exclusion reach for invisible decays of the Higgs boson, using the process $e^+e^-\to HZ$, and considering only the decay channels $Z\to e^+e^-$ and $Z\to \mu^+\mu^-$. The analysis is based on a template fit, in which the signal and background distributions are assumed to be well known from accurate detector simulations and studies of data control samples. 
The results are summarized in Table~\ref{tab:conclusions}. It is  worth mentioning that it is possible to improve these results  by including hadronic Z decays in the analysis.


\begin{table}[htb]
\caption{Discovery and exclusion reach for invisible decays of the Higgs boson for 3.5ab$^{-1}$ of data acquisition at FCC-ee, using the process $e^+e^-\to HZ$, and considering only the decay channels $Z\to e^+e^-$ and $Z\to \mu^+\mu^-$.\label{tab:conclusions}}
\centering
\begin{tabular}{ c | c | c }
& $BR_{95\% limit}$ & $BR_{5\sigma}$\\
\hline
CMS-like & $0.92\pm0.32 \%$ & $2.5\pm 0.2\%$ \\
ILD-like & $0.63 \pm 0.22 \%$ & $1.7\pm 0.1\%$ \\
\end{tabular}
\end{table}

The limits of Table~\ref{tab:conclusions} are translated into the expected bound on DM-nucleon cross section within the framework of Higgs-portal models.  The FCC-ee sensitivity projects to an improvement by two orders of magnitude with respect to the LHC bounds currently available and remains competitive with the reach of planned direct detection experiments for DM masses smaller than about 10 GeV.

\section*{Acknowledgement}


We thank M.~Selvaggi for the help provided with the implementation of the ILD-like performances in our study. We thank P.~Janot, A.~Blondel. E. Perez and D.D'Enterria  for useful inputs and discussions on the design of FCC-ee, and the experimental and physical backgrounds to consider in the analysis. In addition, we want to express our gratitude to Laura Baudis for pointing us to the existing projections for future direct dark matter detection experiments.

\appendix

\section{Comparison between CMS-like and ILD-like designs}
\label{appendix_comparison}

The numerical results shown in Sec.~\ref{sec:results} shows that an ILD-like detector design allows to improve by $\approx 50\%$ the results of a CMS-like detector. This improvement has two causes: (i) the better tracking resolution reduces the width of the  $M_{\rm miss}$ signal peak; (ii) the ILD-like reconstruction benefits of a larger efficiency for the lepton reconstruction and identification. 

In this appendix, we discuss briefly the impact of the tracking resolution on the $M_{\rm miss}$ signal distribution. In an ideal situation, one would push for the best possible tracking resolution. On the other hand, in final states like the one considered in this study the experimental resolution also depends on the knowledge of the collision energy. At a high-luminosity \epem collider, beam-beam interactions introduce an energy spread which randomizes the electron and positron momenta. The typical spread is quite small  (0.2\%) and when computing the missing mass with a CMS-like detector this effect is not visible in the missing mass resolution.
However, the energy spread becomes a limiting factor if one pushes the tracking resolution at the high-precision expected for the ILD-like design.

To show the interplay between tracking resolution and energy spread, we compare in Fig.~\ref{fig:energySpread} the $M_{\rm miss}$ distribution in three scenarios: (i) an ILD-like detector taking data at an \epem collider with no energy spread; (ii) the same ILD-like detector taking data at the FCC-ee, collider with energy spread $0.17\%$ per beam, resulting in $0.12\%$ on the total energy; a CMS-like detector, taking data at the FCC-ee including the energy spread.  
As the figure shows, introducing the energy spread in the simulation deteriorates substantially the resolution of the $M_{\rm miss}$ signal peak of an ILD-like detector, as the  FWHM of the peak increases from 100 MeV without energy spread to 500 MeV with the baseline spread. The FWHM of the peak is roughly linear with the energy spread for a change of $\pm 50\%$ with respect to the baseline value.


\begin{figure}[htb]
 \centering
\includegraphics[width=0.5\textwidth]{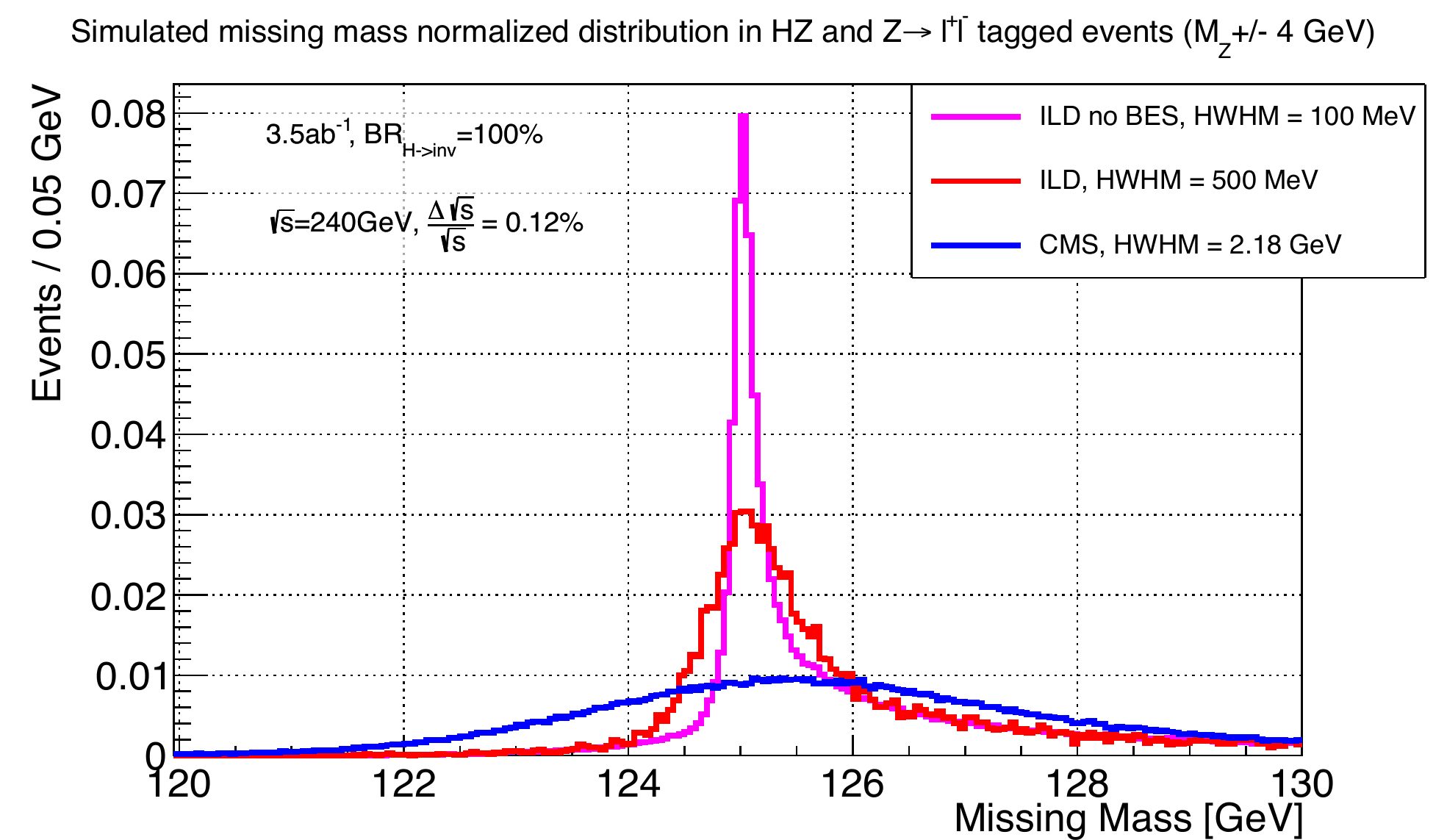}
 \caption{$M_{\rm miss}$ distribution for different configurations of tracking resolution and beam energy spread: (i) an ILD-like detector taking data at an ideal \epem collider with no energy spread; (ii) the same ILD-like detector taking data at the FCC-ee, collider with beam energy spread $0.17\%$, hence $0.12\%$ on the total energy; a CMS-like detector, taking data at the FCC-ee. \label{fig:energySpread}}
\end{figure}


\begin{thebibliography}{99}

\bibitem{ATLASHIGGS}
  G.~Aad {\it et al.} [ATLAS Collaboration],
  Phys.\ Lett.\ B {\bf 716} (2012) 1
  doi:10.1016/j.physletb.2012.08.020
  [arXiv:1207.7214 [hep-ex]].
\bibitem{CMSHIGGS}
  S.~Chatrchyan {\it et al.} [CMS Collaboration],
  Phys.\ Lett.\ B {\bf 716} (2012) 30
  doi:10.1016/j.physletb.2012.08.021
  [arXiv:1207.7235 [hep-ex]].

\bibitem{FCC}
FCC website: http://fcc.web.cern.ch

\bibitem{TLEP}
  M.~Bicer {\it et al.} [TLEP Design Study Working Group Collaboration],
  JHEP {\bf 1401} (2014) 164
  doi:10.1007/JHEP01(2014)164
  [arXiv:1308.6176 [hep-ex]].

\bibitem{FCCeeweb}
FCC-ee website:http://tlep.web.cern.ch/

\bibitem{Oide:2016ipr} 
  K.~Oide {\it et al.},
 ``Design of Beam Optics for the FCC-ee Collider Ring,''
  doi:10.18429/JACoW-IPAC2016-THPOR022
  
 \bibitem{Zimmermann}
  F.~Zimmermann, at KET Workshop on Future \epem Colliders
  https://indico.mpp.mpg.de/event/\\4223/session/1/contribution/4/material/slides/0.pptx


\bibitem{HiggsPortal}
  B.~Patt and F.~Wilczek,
  hep-ph/0605188.

\bibitem{PYTHIA8}
  T.~Sjostrand, S.~Mrenna and P.~Z.~Skands,
  Comput.\ Phys.\ Commun.\  {\bf 178} (2008) 852
  doi:10.1016/j.cpc.2008.01.036
  [arXiv:0710.3820 [hep-ph]].

\bibitem{LEP3}
P.~Azzi {\it et al.},
[arXiv:1208.1662 [hep-ex]]

\bibitem{ATLASHiggstoInv}
  G.~Aad {\it et al.} [ATLAS Collaboration],
  arXiv:1509.00672 [hep-ex].

\bibitem{CMSHiggstoInv}
  S.~Chatrchyan {\it et al.} [CMS Collaboration],
  Eur.\ Phys.\ J.\ C {\bf 74} (2014) 2980
  doi:10.1140/epjc/s10052-014-2980-6
  [arXiv:1404.1344 [hep-ex]].

\bibitem{GiaStrumia}
  A.~De Simone, G.~F.~Giudice and A.~Strumia,
  JHEP {\bf 1406} (2014) 081
  doi:10.1007/JHEP06(2014)081
  [arXiv:1402.6287 [hep-ph]].
  
  \bibitem{ILCPROJECT}
 ILC website: http://www.linearcollider.org/ILC/What-is-the-ILC/The-project
  
  \bibitem{CEPC}
CEPC website: http://cepc.ihep.ac.cn

 \bibitem{HANLIU} 
  T.~Han, Z.~Liu and J.~Sayre,
  Phys.\ Rev.\ D {\bf 89}, no. 11, 113006 (2014)
  doi:10.1103/PhysRevD.89.113006
  [arXiv:1311.7155 [hep-ph]].

\bibitem{CEPCTDR} 
  CEPC-SPPC Study Group,
  IHEP-CEPC-DR-2015-01, IHEP-TH-2015-01, HEP-EP-2015-01.

  
  \bibitem{ILCTDR}
  T.Behnke {\it et al.} {ILC Collaboration}
  The International Linear Collider Technical Design Report - Volume 4: Detectors
  [arXiv:1306.6329 [hep-ex]].
 

\bibitem{DELPHES}
  J.~de Favereau {\it et al.} [DELPHES 3 Collaboration],
  JHEP {\bf 1402} (2014) 057
  doi:10.1007/JHEP02(2014)057
  [arXiv:1307.6346 [hep-ex]].
  
\bibitem{CMSDELPHES}
Cards provided with the software distribution \verb+delphes_card_CMS.tcl+ S.

\bibitem{ILDDELPHES}
Cards provided by the DELPHES collaboration , based on \cite{ILCTDR}.

  
\bibitem{JetAlgo} 
  M.~Cacciari, G.~P.~Salam and G.~Soyez,
  JHEP {\bf 0804} (2008) 063
  doi:10.1088/1126-6708/2008/04/063
  [arXiv:0802.1189 [hep-ph]].

\bibitem{FASTJET}
  M.~Cacciari, G.~P.~Salam and G.~Soyez,
  Eur.\ Phys.\ J.\ C {\bf 72} (2012) 1896
  doi:10.1140/epjc/s10052-012-1896-2
  [arXiv:1111.6097 [hep-ph]].

\bibitem{Angloher:2011uu}
  G.~Angloher {\it et al.},
  Eur.\ Phys.\ J.\ C {\bf 72} (2012) 1971\\
  doi:10.1140/epjc/s10052-012-1971-8
  [arXiv:1109.0702 [astro-ph.CO]].

\bibitem{Angle:2011th}
  J.~Angle {\it et al.} [XENON10 Collaboration],
  Phys.\ Rev.\ Lett.\  {\bf 107} (2011) 051301
   [Phys.\ Rev.\ Lett.\  {\bf 110} (2013) 249901]
  doi:10.1103/PhysRevLett.110.249901, 10.1103/PhysRevLett.107.051301
  [arXiv:1104.3088 [astro-ph.CO]].

\bibitem{Bernabei:2008yi}
  R.~Bernabei {\it et al.} [DAMA Collaboration],
  Eur.\ Phys.\ J.\ C {\bf 56} (2008) 333
  doi:10.1140/epjc/s10052-008-0662-y
  [arXiv:0804.2741 [astro-ph]].

\bibitem{Savage:2008er}
  C.~Savage, G.~Gelmini, P.~Gondolo and K.~Freese,
  JCAP {\bf 0904} (2009) 010
  doi:10.1088/1475-7516/2009/04/010
  [arXiv:0808.3607 [astro-ph]].

\bibitem{Aalseth:2012if}
  C.~E.~Aalseth {\it et al.} [CoGeNT Collaboration],
  Phys.\ Rev.\ D {\bf 88} (2013) 012002
  doi:10.1103/PhysRevD.88.012002
  [arXiv:1208.5737 [astro-ph.CO]].

\bibitem{Agnese:2013rvf}
  R.~Agnese {\it et al.} [CDMS Collaboration],
  Phys.\ Rev.\ Lett.\  {\bf 111} (2013) 25,  251301
  doi:10.1103/PhysRevLett.111.251301
  [arXiv:1304.4279 [hep-ex]].

\bibitem{Behnke:2012ys}
  E.~Behnke {\it et al.} [COUPP Collaboration],
  Phys.\ Rev.\ D {\bf 86} (2012) 5,  052001
   [Phys.\ Rev.\ D {\bf 90} (2014) 7,  079902]
  doi:10.1103/PhysRevD.86.052001, 10.1103/PhysRevD.90.079902
  [arXiv:1204.3094 [astro-ph.CO]].

\bibitem{Akerib:2013tjd}
  D.~S.~Akerib {\it et al.} [LUX Collaboration],
  Phys.\ Rev.\ Lett.\  {\bf 112} (2014) 091303
  doi:10.1103/PhysRevLett.112.091303
  [arXiv:1310.8214 [astro-ph.CO]].


\bibitem{Djouadi2011}
Abdelhak Djouadia, Oleg Lebedev, Yann Mambrini and Jeremie Quevillon,
arXiv:1112.3299v3 [hep-ph] 

\bibitem{fN_LQCD} 
  R.~D.~Young and A.~W.~Thomas,
  Phys.\ Rev.\ D {\bf 81} (2010) 014503
  doi:10.1103/PhysRevD.81.014503
  [arXiv:0901.3310 [hep-lat]].

\bibitem{MILC} 
D.~Toussaint {\it et al.} [MILC Collaboration],
  Phys.\ Rev.\ Lett.\  {\bf 103} (2009) 122002
  doi:10.1103/PhysRevLett.103.122002
  [arXiv:0905.2432 [hep-lat]].

\bibitem{HLLHC}
CMS Collaboration, 
  [arXiv:1307.7135v2 [hep-ex]].

\bibitem{reviewDM}
  Laura Baudis,
  [arXiv:1509.00869 [ astro-ph]].
\end{thebibliography}
\end{document}